# Aharonov-Bohm resistance magneto-oscillations on single-nanohole graphite and graphene structures

Yu.I. Latyshev[1], A.P. Orlov[1], V.A. Volkov[1], V.V. Enaldiev[1], I.V. Zagorodnev[1],
O.F. Vyvenko[2], Yu.V. Petrov[2], P. Monceau[3]

[1]V.A. Kotelnikov Institute of Radio-Engineering and Electronics of RAS, Mokhovaya 11-7, 125009 Moscow, Russia,
[2]IRC for Nanotechnology of St. Petersburg State University, Uljanovskaya 1, Petrodvorets, 198504 St. Petersburg, Russia
[3]Institut Neel, CNRS/UJF UPR 2940, 25 rue des Martyrs, BP 166, 38042 Grenoble, cedex 9, France

**Graphene is a stable single atomic layer material exhibiting two-dimensional electron gas of massless Dirac fermions of high mobility. One of the intriguing properties of graphene is a possibility of realization of the Tamm-type edge states. These states differ from the usual surface states caused by defects, impurities and other imperfections at the edge of the system, as well as they differ from the magnetic edge states caused by skipping cyclotron orbits. The Tamm states result from breaking of periodic crystal potential at the edge, they can exist even at zero magnetic field and form a conducting band. Until recently those states have been observed in graphene only by local STM technique and there were no direct experiments on their contribution to transport measurements. Here we present the experiments on Aharonov-Bohm (AB) oscillations of resistance in a single-nanohole graphite and graphene structures, it indicates the presence of conducting edge states cycling around nanohole. An estimation show the penetration depth of the edge states to be as short as about 2 nm. The oscillations persist up to temperature T=115 K and the T-range of their existence increases with a decrease of the nanohole diameter. The proposed mechanism of the AB oscillations based on the resonant intervalley backscattering of the Dirac fermions by the nanohole via the Tamm states. The experimental results are consistent with such a scenario. Our findings show a way towards interference devices operating at high temperatures on the edge states in graphene.**

There are two types of quantum oscillations of the electrical resistance of common 2D conducting structures under magnetic field $H$: 1) magneto-oscillations periodic in $1/H$, due to the Shubnikov-de Haas effect, and 2) magneto-oscillations periodic in $H$ due to the Aharonov-Bohm (AB)-type effect. $H$-periodic oscillations usually observed on samples of narrow mesoscopic rings, the period of oscillations being defined by the commensurability of the magnetic flux through the ring and the magnetic flux quantum. We constructed a system of non-standard geometry, showing the AB effect in electron transport. This is the structure of graphene-on-graphite with a single nanohole.

The principal possibility to observe the AB effect in the samples of non-ring geometry is related with the existence of edge states cycling the nanohole perimeter similarly as it was observed on the surface of nanowires of topological insulator $Bi_2Se_3$[1]. The origin of conductive surface states in topological insulators is considered to be related with topological reasons[2], while graphene is not topological insulator. However, a theory[3-5] predicts the existence of conductive edge states at the edge of graphene as well. The reason for this is a sharp break of the crystal potential at the edge of graphene. In fact, this is a modification of the Tamm states[6] in graphene. Some evidence of the edge states has been demonstrated in graphene by STM[7,8].

The presence of the conducting edge states in graphene in principle could be checked by an observation of the AB effect on a single nanohole structure. However, until recently the AB effect in graphene has been observed only on ring shaped samples[9-17] or on the lattice of nanohole[18] where effect on single hole has been masked by a presence of the lattice. Some indication of the AB effect on a single nanohole has been found in experiments on nanothin (50 nm-thick) graphite crystals irradiated with heavy ions[19]. Low dose irradiated samples with randomly distributed columnar defects exhibited field-periodic oscillations of resistance with a period of 7.5 T roughly correspondent to the flux quantum per columnar defect area. It was also shown that the period of oscillation does not depend on graphite thickness down to thickness of 1 nm[20]. That pointed to the significant contribution of the surface graphene layer to the quantum oscillations of nanothin graphite. This result is consistent with recent STM[21], cyclotron resonance[22] and Raman spectroscopy[23,24] experiments showed that the surface layer of graphite is often represented as a graphene layer of an exceptional quality. This graphene-on-graphite system is the most attractive for thin enough graphite samples to avoid the shunting of the surface graphene layers by the bulk. Here we report on the observation of the AB effect on the graphene and graphite structures with a single nanohole.

Large area flakes of natural graphite with thickness down to 30 nm were obtained by cleavage from the large crystal graphite using adhesive tape, followed by dissolving the adhesive layer in acetone. At the second stage the crystal can be thinned by soft plasma etching down to the atomic thickness of about 1 nm[25]. The scanning Raman spectroscopy showed high uniformity of the thinned crystals over the lateral size of hundreds of microns. For comparison we have also used commercial graphene samples.

The samples have been processed to form Hall bar geometry. The nanoholes have been introduced by three independent techniques (1) by irradiation with heavy ions[19] (2) by FIB technique using Ga-ions, (3) by helium ion microscope using He-ions. The irradiation has been done with Xe-ions with energy of 100 Mev at Large Heavy Ion National Accelerator (GANIL), Caen, France. The estimated from AFM images diameter of columnar defects was $d = 24$ nm[19] (Fig. 1a). We used FIB technique for fabrication small holes with diameter down to $d = 35$ nm (Fig. 1 b). To produce the smallest hole with diameter $d = 20$ nm we used helium ion microscope (Fig. 1c, d).

Results of resistance measurements are shown on Fig. 2 for a structure on thin graphite with a single nanohole produced by FIB (Fig. 2a) and graphene structure with a single nanohole produced by helium ion microscope (Fig. 2b). The common feature for both structures is the presence of field-periodic oscillations at high magnetic fields.

In low fields the Shubnikov - de Haas oscillations are also clearly observed on thin graphite samples. They have inversed field periodicity with a period of 0.2 T$^{-1}$ and are terminated at fields above $\approx 8$T[19], when an energy of the first Landau level exceeds Fermi energy.

We compared the period of oscillations for three single nanohole samples of different diameters (see Table 1). Sample #1, shown on Fig. 1 c, has been fabricated on graphene. Sample #4 contained columnar defects.

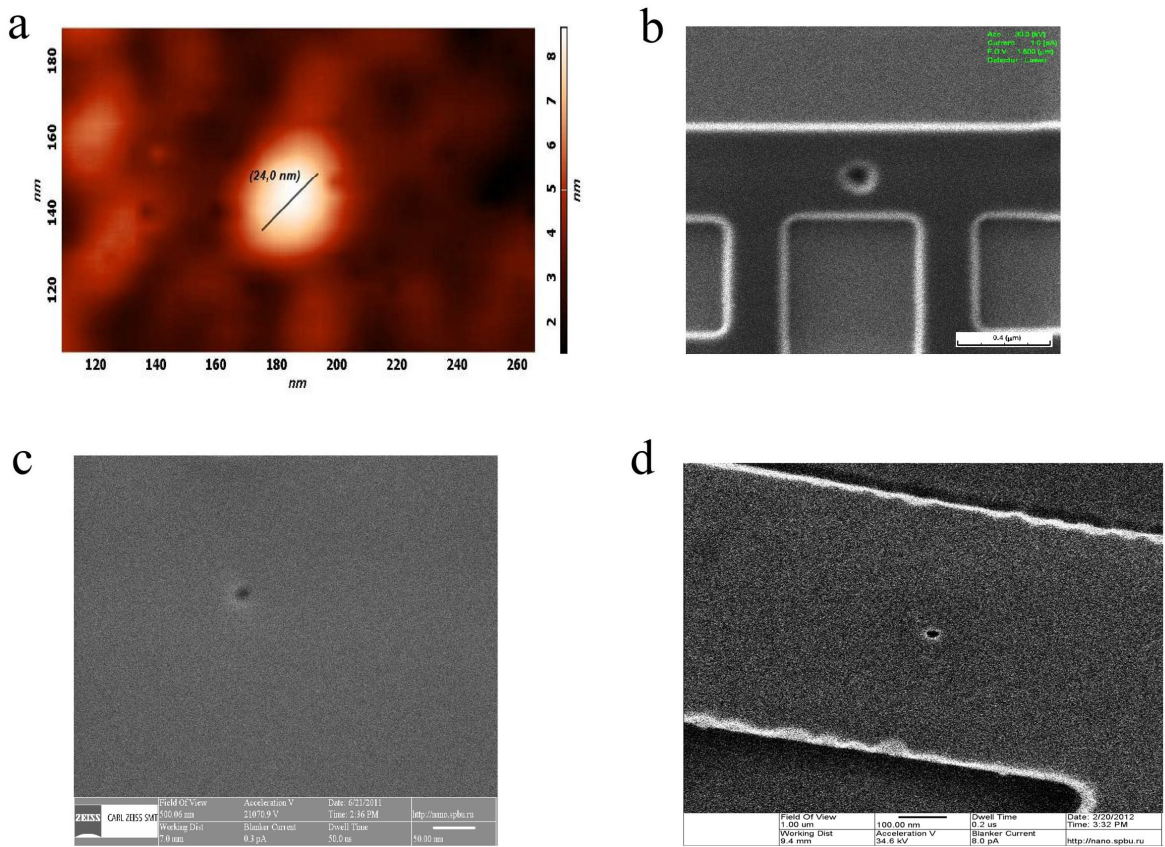

**Figure 1 | Experimental realization of graphene nanohole structures and results.** *a, Single holes produced by heavy ion irradiation (AFM image), b, by FIB (SEM image) and c, by helium ion microscope (SHIM image) on graphene (c) and thin graphite (a, b, d).*

Within experimental uncertainty of about 10% we found that the period of oscillations $\Delta H$ for all samples corresponds to the flux quantization in a hole

$$\Delta H \pi D^2 / 4 = \Phi_0 \qquad (1)$$

with $\Phi_0 = hc/e$ – the flux quantum, $D$ – a diameter of the hole, if one consider that the main contribution to the effect comes from the carriers which orbits localized very near to the edge of the hole. This periodicity was observed in ring shaped samples of graphene[9]. We can therefore attribute oscillations to the quantum interference of massless Dirac fermions (DFs) in a band of conducting edge states. That was also supported by previous observations[20], where the period of magneto-oscillations in the irradiated samples of thin graphite did not depend on their thickness down to the thickness of 1 nm.

**Table 1.** *The table includes the following samples: #1 – graphene structure with a hole produced by HIM, #2 –thin graphite structure with columnar defects, #3 – thin graphite structure with a hole produced by FIB, #4– thin graphite structure with a hole produced by HIM. The thickness of thin graphite structures have been varied within 30-50 nm. Parameter $D_{eff}$ has been calculated using Eq. (1).*

| Sample No | $\Delta H$, T | $D_{geom.}$, nm | $D_{eff.}$, nm | $(D_{geom} - D_{eff})/2$, nm |
|---|---|---|---|---|
| #1 | 9.0 | 20±1 | 24±0.1 | 2.0±0.5 |
| #2 | 7.5 | 24±1 | 27±0.1 | 1.5±0.5 |
| #3 | 3.2 | 37±2 | 41±0.2 | 2.0±1.0 |
| #4 | 6.0 | 25±1 | 30±0.2 | 2.5±0.5 |

Another point to discuss is why the orbits of the DFs contributed to the oscillations are so close to the edge. Actually, all the possible orbits of DFs around the hole (for example, skipping cyclotron orbits) should neglect interference effect except the possibility of the existence of the edge states. As for the skipping along the hole perimeter orbits, they are very sensitive to the edge potential profile, in contrast to the Tamm orbits. The Tamm edge states can play the role of the ring keeping carriers near the edge of the hole. With the experimental accuracy we can estimate the effective width of this type effective ring. Table 1 shows comparison of the geometrical size of the hole and the size of the effective 1D ring producing the same effect. One can see that the radius of the effective AB-ring is always higher the geometrical one and the difference is about 2 nm with experimental accuracy. That gives an experimental estimation of the width of the effective ring associated with the edge states, i.e. the penetration depth of the edge states.

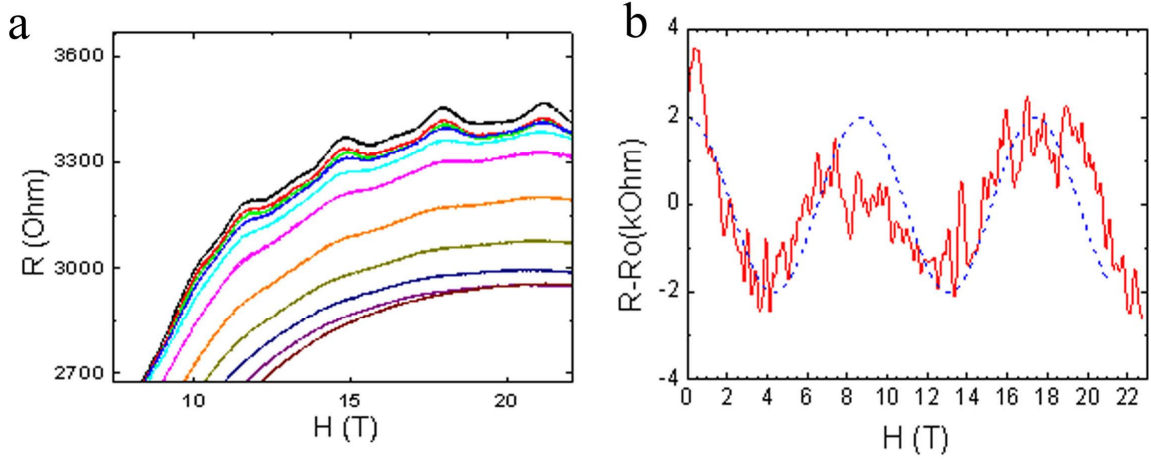

**Figure 2| Aharonov-Bohm resistance magneto-oscillations. *a.** Field-periodic resistance oscillations for thin graphite single hole structure with FIB made nanohole with D = 37 nm (T=1.5, 4.2, 10, 15, 15, 20, 30, 45K from top to bottom); **b**, graphene structure with a single nanohole made by helium ion microscope, D = 20 nm, T=4.2 K.*

We discuss now a temperature dependence of the amplitude of oscillations. At the first experiments on graphene rings the AB oscillations have been observed only below liquid helium temperature range. The effect on nanoholes persists to much more higher temperature.

We consider now the less noisy experiment on thin graphite single-hole-structure. Fig. 3a shows oscillating part of magnetoresistance extracted from the data of Fig. 2a by subtracting of monotonic part. One can see that oscillations persist up to the temperatures as high as 50 K. The four main peaks marked by the upper arrows are clearly seen at fields above 10 T. Their spacing $\Delta H$ corresponds to the flux quantum per nanohole area, following Eq. 1.

Fig. 3b shows the temperature dependence of the height of one of these peaks $A$ observed at $H = 18$ T. That clearly shows exponential dependence $A \propto \exp(T/T_0)$ with $T_0 \approx 17$ K. Weak T-damping of oscillations is consistent with a theory of the edge states discussed below.

The next important point revealed from the experiment is the existence of relatively small peaks marked at the Fig. 3a by the upward arrows. Two series of peaks can be considered as shifted by $\pi$ series with the same periodicity of $\Phi_0$ or as series of oscillations with periodicities of $\Phi_0$ and $\Phi_0/2$. We can extract temperature dependence of these type oscillations only from the peak at $H = 10$ T where that is more or less clearly resolved. The comparison of temperature dependence of peaks of those two different series shows the same exponential dependence indicating their common origin that originates from the edge states.

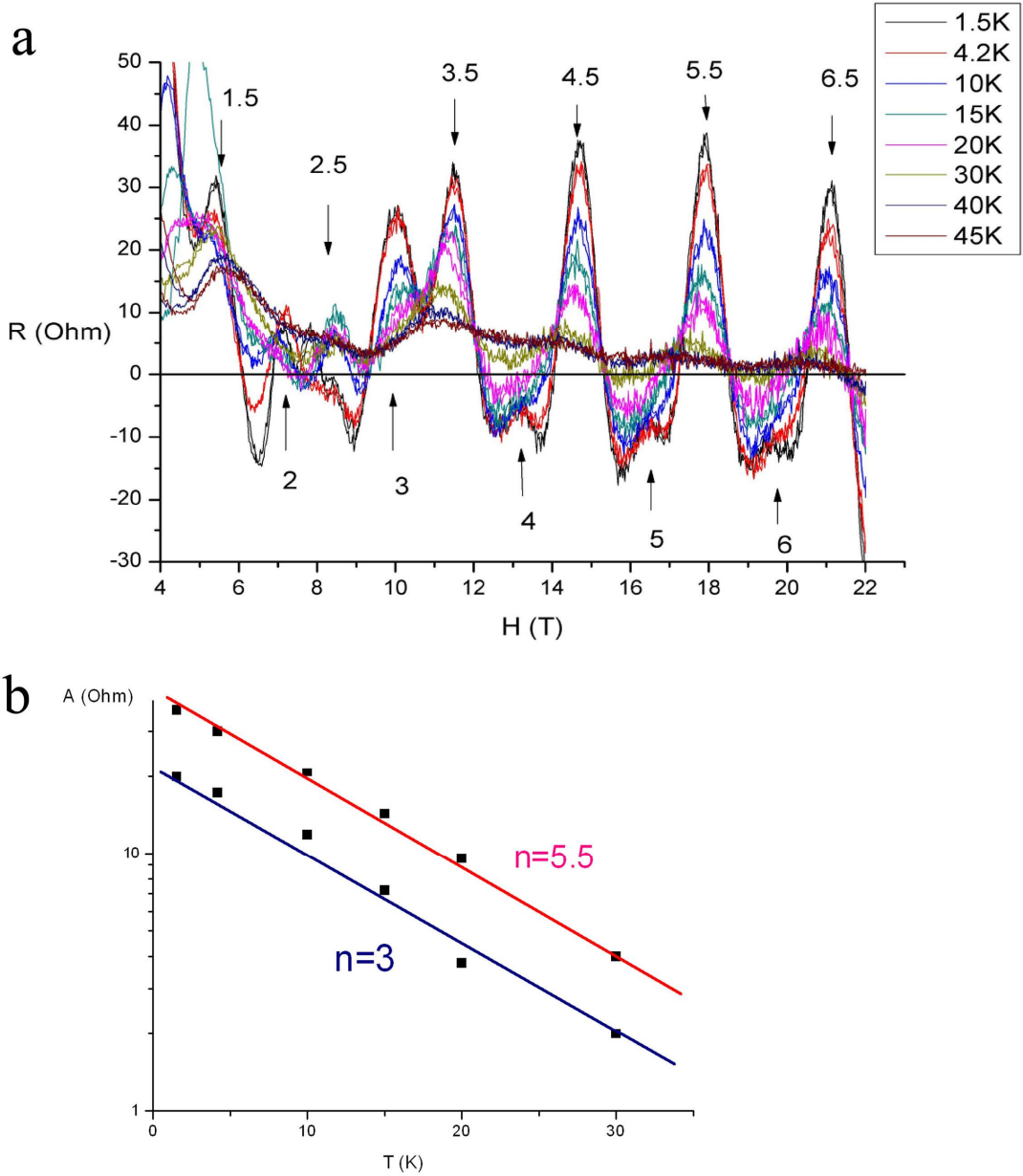

**Figure 3 | Temperature influence on magneto-oscillations.** *a, Oscillating part of magnetoresistance of structure #2 at various temperatures. The downward arrows show main series, corresponding to* $\Phi_\downarrow = n\Phi_0 + 1/2$ *(n is an integer), while upward arrows mark an additional series* $\Phi_\uparrow = n\Phi_0$. *b, Temperature dependences of the amplitude of resistance oscillations of sample #2 for* $\Phi/\Phi_0 = 5.5$ *(red line) and* $\Phi/\Phi_0 = 3$ *(blue line).*

One can consider that characteristic temperature $T_0$ is related with the typical energy of the edge state as $kT_0 = E_0 = \hbar v_0 / R$. That gives an estimate for $v_0$, $v_0 = 5 \cdot 10^6$ cm/s. To compare temperature dependences for samples #3 and #4 we found that for nanohole of smaller diameter $T_0$ value increases roughly proportional to $1/R_{eff}$.

**Theory and comparison with experiment**

Remarkable band structure of graphene consists of two almost independent pair of cones, that are often called valleys. Let us paint them red and blue for convenience. Electronic properties of graphene are modified near an edge so that Tamm-type edge state can appear. The wave function of carriers at the edge state is exponentially localized near the interface. Neglecting the intervalley scattering one can characterize the edge by a real phenomenological parameter $a$ [4]. The value of $a$ can be determined by a comparison with experiment. However, a comparison with the microscopic model calculations [5] shows that it is small and positive ($0 < a \ll 1$) and below we will assume that for simplicity. For a half-plane graphene sample the spectrum of the edge Tamm states in zero magnetic field is (Fig. 4a):

$$E_\tau(k_\parallel) = \tau 2av\hbar k_\parallel, \quad \tau k_\parallel \geq 0. \tag{2}$$

Here $k_\parallel$ is one-dimensional electron momentum along the edge, measured from the center of the valleys, $v = 10^8$ cm/s is the effective speed of light in the bulk graphene spectrum, the index $\tau = \pm 1$ enumerates the valleys in graphene. It is important that the sign of $k_\parallel$ is determined by the valley number $\tau$. For small $a$ localization length of the edge state at the Fermi level is $x_T = 1/|k_r| = 1/|k_b|$. The comparison with experiment, Table 1, gives the estimate $x_T = 2$ nm.

Edge states also exist at the edge of a round hole ("antidot") with radius $R$ in an infinite sheet of graphene as well. Electrons trapped in the edge states, behave like electrons in a narrow ring ($x_T \ll R$), which allows to expect manifestations of the Aharonov-Bohm effect. A finite perimeter of the antidot leads to a quantization of the parallel to the edge momentum component. Discrete edge states now are characterized by half-integer total angular momentum $j_\tau$ ($\tau j_\tau > 0$). In the semiclassical approach, the spectrum of these edge states is obtained from (2) substituting $k_\parallel$ for $j_\tau/R$. Although these states in the absence of a magnetic field are quasistationary, their finite lifetime (due to decay into bulk states) is a large value in the actual case of small $a$.

In a magnetic field $H$ the spectrum of edge Tamm states in antidot has semiclassical asymptotic at $a \ll 1$:

$$E_\tau = \tau 2a \frac{\hbar v}{R}(j_\tau + \Phi/\Phi_0 - \tau/2), \tag{3}$$

here $\Phi = H\pi R^2$, $\Phi_0 = hc/e$ and now $j_\tau$ satisfy $\tau(j_\tau + \Phi/\Phi_0) > 0$.

This asymptotic behavior is the more justified than less magnetic field affects the orbital part of the wave function, i.e. when the localization length of the edge state is small to compare with the magnetic length. Under conditions of our experiment that is valid for all magnetic fields. Knowing $x_T = 2$ nm and a value of field corresponding to the last Shubnikov' oscillation, we can estimate the parameter $a$ as $a = 0.05$.

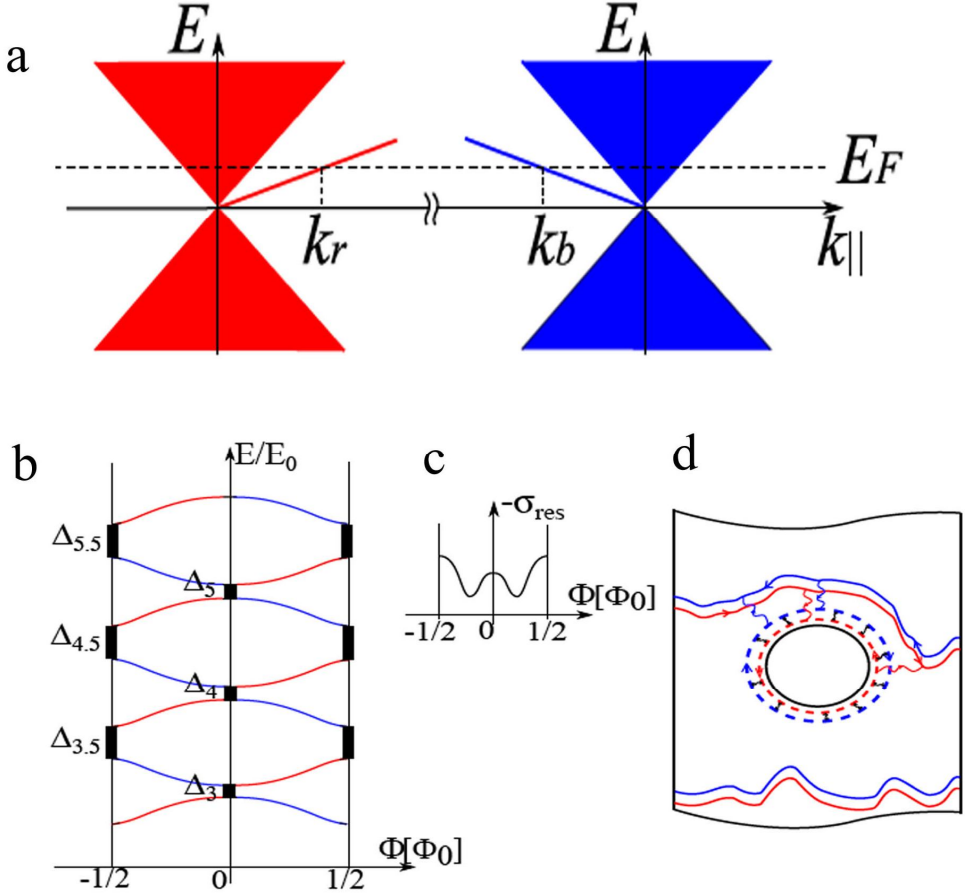

**Figure 4| Edge states around graphene nanohole. a,** The red and the blue rays are the edge Tamm states contra-propagating along the graphene semi-plane. There are two Tamm states at the Fermi level: $k_r$ in the red valley with positive velocity and $k_b$ in the blue valley with negative velocity. Bulk continuum states are shadowed. **b,** Spectrum of the edge Tamm states in antidot for different $j_+, j_-$ as a function of the magnetic flux passing through the antidot area. The spectrum has a band character, and the flow through the antidot plays a role of quasi-momentum in the reduced zone scheme. The red (blue) color corresponds to the valley with $\tau = +1$ ($\tau = -1$), $E_0 \approx 2a\hbar v/R$. The red-blue scattering results in band gaps (vertical bold lines). Gaps are formed by anticrossing of red and blue edge states with angular momenta $j_+$ and $j_-$. Gap values are denoted by index $j = (j_+ + j_-)/2$. **c,** Intervalley contribution in conductivity in the reduced zone scheme. Peaks correspond to resonant red-blue back-scattering. Two series of the peaks are connected with passing of the magnetic flux through the centre and boundary of zone shown on **a**. **d,** Contra–propagating trajectories of the orbit centers for different valleys for the main Landau level with $N = 0$ in a smooth impurity potential in the real space. One of them is close to the antidote and can experience back scattering as a result of intervalley scattering at some values of the trapped flux.

Electron in the edge state moves periodically around the antidot (clockwise in one valley and against it in the other), picking up additional Aharonov-Bohm phase in a magnetic field. Its wave function satisfies the Bloch theorem, and the spectrum of the edge states has a band character, while the magnetic flux plays the role of an effective quasi-momentum. Intravalley scattering does not essentially change this picture. Another situation happens when one takes into account a weak intervalley scattering. The potential of this scattering is similar to a periodic potential, and leads to the formation of energy gaps in the band spectrum in Fig. 4b, if the conditions of intervalley ("blue-red") resonance are fulfilled:

$$\frac{\Phi}{\Phi_0} = \frac{j_+ + j_-}{2} \qquad (4)$$

Mode of this resonance, corresponding to the gaps in the spectrum of Fig. 4b, leads to a strong backscattering, which, in turn, according to Landauer formula leads to the negative peaks in the conductance. Consider that in the magnetic quantum limit when the AB oscillations are observed, the Fermi level of the system is close to the Dirac point, which fluctuates strongly in space due to the formation of "puddles" of electrons and holes[26]. If the spatial scale of these fluctuations is comparable with the size of the antidot, and the energy scale is comparable with the energy of the quantization of the edge states, the resonance (4) does not depend on the position of the Fermi energy and its temperature smearing. This qualitatively explains the weak temperature dependence of the observed AB oscillations.

As shown above, for a velocity of the edge fermions extracted from the experimental temperature dependence of the amplitude of oscillations is 20 times less than for bulk ones. That may be explained if to consider the parameter $a = 0.05$. That is consistent with an independent estimate of $a$ made above.

With the magnetic field variation the energy of Tamm states in the red valley $E_+(\Phi, j_+)$ periodically coincide with Tamm levels of the blue valley $E_-(\Phi, j_-)$ with a periodicity of $\Phi_0/2$. Because of $j$ takes only half-integer values from (4) it follows that $\Phi/\Phi_0$ can accept either half-integer or integer values. The comparison with experiment shows that the first condition is responsible for the main series of oscillations, while the second - for the complementary series.

**Conclusions**

We found that graphene-on-graphite and graphene nanostructures containing single nanoholes (antidotes) with diameter of 20-40 nm show field-periodic resistance oscillations with flux periodicity close to flux quantum per nanohole area. The result is considered as the Aharonov-Bohm effect on the conducting edge states of Tamm-type localized near the edge of hole. From the experiment we have got an estimate of the values of penetration depth of the edge states and velocity of carriers occupied those states. The experimental results are consistent with a theory of the edge states.

**Acknowledgments**

The work has been supported by RFBR grants No 11-02-01379-a, No 11-02-01290-a, No 11-02-121687-ofi-m-2011, by grant of the Russian Ministry of Education and Science, agreement No 8033, by programs of RAS, by the European Comission from the 7[th] framework programme "Transnational Access", contract N0 228043-Euromagnet II- Integrated Activities.

**References**

1. Peng H., Lai K., Kong D., Meister S., Chen Y., Qi X.-L., Zhang S.-C., Shen Z.-X., Cui. Y. Aharonov–Bohm interference in topological insulator nanoribbons. *Nature Mater.* **9**, 225 (2010).

2. Hasan M. Z., Kane C. L.. Colloquium: Topological insulator. *Rev. Mod. Phys.* **82**, 3045 (2010).

3. Nakada K., Fujita M., Dresselhaus M. S. The edge state in graphene ribbons: Nanometer size effects and edge shape dependence. *Phys. Rev. B.* **54**, 17954-17961 (1996).

4. Volkov V. A., and Zagorodnev I. V. Electrons near a graphene edge. *Low Temp. Phys.* **35**, 2-5 (2009).

5. van Ostaay J. A. M, Akhmerov A. R., Beenakker C. W. J., and Wimmer M.. Dirac boundary condition at the reconstructed zigzag edge of graphene. *Phys.Rev.B* **84**, 195434 (2011).

6. Tamm I.E.. On the possible bound states of electrons on a crystal surface. *Phys. Z. Soviet Union* **1**, 733 (1932).

7. Ritter K. A., and Lyding J. W. The influence of edge structure on the electronic properties of graphene quantum dots and nanoribbons. *Nature Mater.* **8**, 235-242 (2009).

8. Tao C., Jiao L., Yazyev O.V., Chen Y.-C., Feng J., Zhang X., Capaz R. B., Tour J. M., Zettl A., Louie S. G., Dai H., and Crommie M.F. Spatially resolving edge states of chiral graphene nanoribbons. *Nature Phys.* **7**, 616-620 (2011).

9. Russo S., Oostinga J. B., Wehenkel D., Heershe H. B., Shams Sobhani S, Vandersypen L. M. K., and Morpurgo A. F. Observation of Aharonov-Bohm conductance oscillations in a graphene ring. *Phys. Rev. B*. **77**, 235404 (2007).

10. Huefner M., Molitor F., Jacobsen A., Pioda A., Stampfer C., Ensslin K., and Ihn T. Investigation of the Aharonov-Bohm effect in a gated graphene ring. *Phys. Status Solidi B*, **245**, 2756-2759 (2009).

11. Wurm J., Wimmer M., Baranger H. U., and Richter K. Graphene rings in magnetic fields: Aharonov-Bohm effect and valley splitting. *Semicond. Sci Technol.* **25**, 034003 (2010).

12. Huefner M., Molitor F., Jacobsen A., Pioda A., Stampfer C., Ensslin K., and Ihn T. The Aharonov-Bohm effect in a side gated graphene ring. *New. J. Phys.* **12**, 043054 (2010).

13. Weng L., Zhang L., Chen Y.,P., and Rokhinson L. P. Atomic force microscope local oxidation nanolithography of graphene. *Appl. Phys. Lett.* **93**, 093107 (2008).

14. Yoo J. S., Park Y. W., Skakalova V., and Roth. Shubnikov-de Haas and Aharonov-Bohm effects in a graphene nanoring structure. *Appl. Phys. Lett.* **96**, 123112 (2010).

15. Smirnov D., Schmidt H., and Haug R. G. Aharonov-Bohm effect in an electron-hole graphene ring system. *Appl. Phys. Lett.* **100**, 203114 (2012).

16. Nam Y., Yoo S.Y, Park Y. W. Lindvall N., Bauch T., and Yurgens A. The Aharonov-Bohm effect in graphene rings with metall mirrors. *Carbon.* **50**, 5562-5568 (2012).

17. *for review see:* Schelter J., Recher P., Trauzettel B. The Aharonov-Bohm effect in graphene rings. *Sol. State Commun*. **152**, 1411-1419 (2012).

18. Shen T., Wu Y. Q., Capano M. A., Rokhinson L. P., Engel L.W., and Ye P. D. Magnetoconductance oscillations in graphene antidot array. *Appl. Phys. Lett*. **93**, 122102 (2008).

19. Latyshev Yu. I., Latyshev A. Yu., Orlov A. P., Schekin A.A., Bykov V. A., Monceau P., van der Beek C. J., Konczykowski M., and Monnet I. Field-periodic magnetoresistance oscillations in thin graphite single crystals with columnar defects. *JETP Lett.* **90**, 480-484 (2009).

20. Latyshev Yu. I., Orlov A. P., Shustin E. G., Isaev N. V., Escoffier W., Monceau P., van der Beek C. J., Konczykowski M., Monnet I. Aharonov-Bohm effect on columnar defects in thin graphite and graphene. *J. Phys.:Conf. Ser.* **248**, 012001 (2010).

21. Li G., Luicann N., and Andrei E. Y. Scanning Tunneling Spectroscopy of Graphene on Graphite. *Phys. Rev. Lett.* **102,** 176804 (2009).

22. Neugebauer P., Orlita M., Faugeras C., Barra A.-L, and Potemski M. How Perfect Can Graphene Be? *Phys. Rev. Lett.* **103**, 136403 (2009).

23. Faugeras C., Amado M., Kossacki P., Orlita M., Kuehne M., Nicolet A.A., Latyshev Yu.I., and Potemski M. Magneto-Raman Scattering of Graphene on Graphite: Electron Scattering and Phonon Excitations. *Phys. Rev. Lett.* **107**, 036807 (2011).

24. Kuhne M., Faugeras C., Kossacki P.,. Nicolet A. A. L, Orlita M., Latyshev Yu.I., and Potemski M. Polarization-resolved magneto-Raman scattering of graphenelike domains on natural graphite. *Phys. Rev. B*. **85**, 195406 (2012).

25. Latyshev Yu. I., Orlov A. P., Peskov V. V., Shustin E. G., Schekin A. A., Bykov V. A. Graphene production by etching natural graphite single crystals in a plasma-chemical reactor based on beam-plasma discharge. *Doklady Physics*. **57**, 1-3 (2012).

26. S. Das Sarma, S. Adam, E. Hwang, E. Rossi. Electronic transport in two-dimensional graphene. *Rev. Mod. Phys.* **83**, 407 (2011).